# WFMOS::Sounding the Dark Cosmos

Bruce A. Bassett[1,3], Bob Nichol[1,4], Daniel J. Eisenstein[2,5] and the **WFMOS Feasibility Study Dark Energy Team**[6]

[1] *Institute of Cosmology and Gravitation, University of Portsmouth, PO12EG, UK*
[2] *Steward Observatory, University of Arizona, 933 N. Cherry Ave., Tucson, AZ 85121*

[3] bruce.bassett@port.ac.uk
[4] bob.nichol@port.ac.uk
[5] deisenstein@as.arizona.edu

[6] Full WFMOS science team list available at http://www.dsg.port.ac.uk/~bruce/kaos/team.html See also http://www.noao.edu/kaos/.

**ABSTRACT**

Vast sound waves traveling through the relativistic plasma during the first million years of the universe imprint a preferred scale in the density of matter. We now have the ability to detect this characteristic fingerprint in the clustering of galaxies at various redshifts and use it to measure the acceleration of the expansion of the Universe. The Wide-Field Multi-Object Spectrograph (WFMOS) would use this test to shed significant light on the true nature of dark energy, the mysterious source of this cosmic acceleration. WFMOS would also revolutionise studies of the kinematics of the Milky Way and provide deep insights into the clustering of galaxies at redshifts up to z~4. In this article we discuss the recent progress in large galaxy redshift surveys and detail how WFMOS will help unravel the mystery of dark energy.

Perhaps the greatest surprise in cosmology since the original discovery of the expansion of the universe in the early 1930's by Edwin Hubble has been the compelling evidence that the cosmic expansion rate has been accelerating rather than slowing down in the past few billion years of cosmic history. Before 1998, models of an accelerating Universe were very unfashionable since assuming that Einstein's jewel – General Relativity – is correct, cosmic acceleration requires the cosmos to be dominated by energy with negative pressure or by Einstein's "greatest blunder" – the cosmological constant, Λ. However, the magnitude of the cosmological constant required to just accelerate the cosmos today is 120 orders of magnitude smaller than the natural quantum gravity scale set by the Planck mass, which would make it the single worst theoretically estimated quantity in the history of science!

However, in 1998, cosmology was turned on its head as observations of distant supernovae by two independent teams showing that these supernovae were fainter and hence more distant than could be explained in a decelerating universe, thereby forcing the cosmology community to confront the mysterious possibility that our Universe is indeed currently accelerating.

Since 1998, the evidence for this acceleration has steadily mounted and improved, yet our knowledge of the underlying physics of the cosmic acceleration has gone almost nowhere. We still do not know whether Einstein's theory of gravity is wrong, whether the acceleration is caused by the cosmological constant or by a completely new form of matter such as the scalar fields often invoked (but as yet undetected) as sources of acceleration during the first few fractions of a second after the Big Bang.

Deciding between these three possible sources of acceleration will be one of the major thrusts in cosmology in the next decade with scores of dedicated surveys and experiments in planning or execution attempting to address the nature of "dark energy", as the mysterious source of acceleration has been dubbed. Currently there are several major new supernovae searches (ESSENCE, SNLS, SDSS-II) which, by 2008, will enhance our understanding of dark energy by measuring the distances to several hundred new Type Ia supernovae (SNIa). SNIa measurements are however fundamentally limited by the fact that we do not physically understand the mechanisms of supernova explosion. In addition there are many possible astrophysical uncertainties, e.g., progenitor bias, absorption of emitted photons by intervening dust and possible redshift evolution of SNIa.

In addition, SNIa distance measurements are not ideal as dark energy discriminators since, given a dark energy model, one must perform two integrals over redshift to derive the corresponding predicted distance (to then compare with the observed distance provided by the SNIa). This double integral smears out any interesting or distinctive fingerprints of the underlying physics implying that one needs a very large number of SNIa to be able to differentiate between different dark energy models. This is exactly the plan with future SNIa surveys such as the Large Synoptic Survey Telescope (LSST), Dark Energy Survey (DES) and the SuperNova Acceleration Probe (SNAP), a satellite mission that would fly after 2014 and would detect nearly 2000 SNIa in the redshift range $0 < z < 1.7$.

An alternative, and highly profitable, approach is to measure the Hubble expansion rate, *as a function of redshift, i.e. H(z)*. This would be advantageous since $H(z)$ is linked to the dark energy models through a single redshift integral only, implying that less smearing occurs and making model discrimination easier. Measuring $H(z)$ is very ambitious however, as consensus for the local (i.e. $z=0$) value, $H_0$, of this rate has only just been reached after decades of heated debate. How can we therefore hope to track the evolution of $H$ with time? We will show that new methods will not only detect $H(z)$, but measure it to an accuracy of less than 3% at certain redshifts, significantly better than we know its value at z=0, i.e. today!

## A rod for dark energy

One way to do exactly this is through the observation of "acoustic oscillations" in the distribution of radiation and matter in the Universe. These acoustic oscillations originate when the cosmos was only about

100,000 years old and was essentially just a smooth hot plasma consisting of photons (i.e. light), electrons, protons, neutrons and dark matter.

The radiation pressure competed with the attractive force of gravity and set up oscillations in the photon fluid. The tight coupling between electrons and photons due to Thompson scattering then caused the baryons to move in unison with the radiation and hence the combined fluid oscillated due to these sound waves causing it to "ring" like a bell. These acoustic oscillations imprint a preferred scale into the distribution of radiation matter corresponding to the sound horizon and were detected convincingly in 2000 through the TOCO, BOOMERanG and MAXIMA balloon observations of the Cosmic Microwave Background (CMB) and mapped over the whole sky by the WMAP satellite in 2003.

Since the baryons and photons were locked in a cosmic dance before the universe became neutral, the oscillations are also imprinted in the spectrum of galaxies we see today. The amplitude of the oscillations in the galaxy clustering is suppressed by an additional factor of $\Omega_b \sim 0.05$, the baryonic contribution to the cosmic energy budget, which makes the oscillations significantly harder to detect in the galaxy distribution than in the CMB. Further, unlike detecting the oscillations in the CMB, which required sufficient angular resolution, detection of the baryon oscillations in the galaxy distribution requires very large survey volume (of order 1 $h^{-3}$ $GPc^3$) to suppress cosmic variance errors and accurate measurements of the redshifts of the galaxies.

It was only in January 2005 that two international teams of cosmologists announced the first convincing detection of these oscillations in the local distribution of galaxies, thus directly demonstrating that the complex clumping of galaxies observed in the present-day Universe evolved directly from these oscillations in the early Universe. This discovery thus provides the "missing link" between the local, matter-dominated cosmos and the young, radiation-dominated, Universe, and was only possible because of two new redshift surveys of galaxies designed and initiated over a decade ago in the early 1990s.

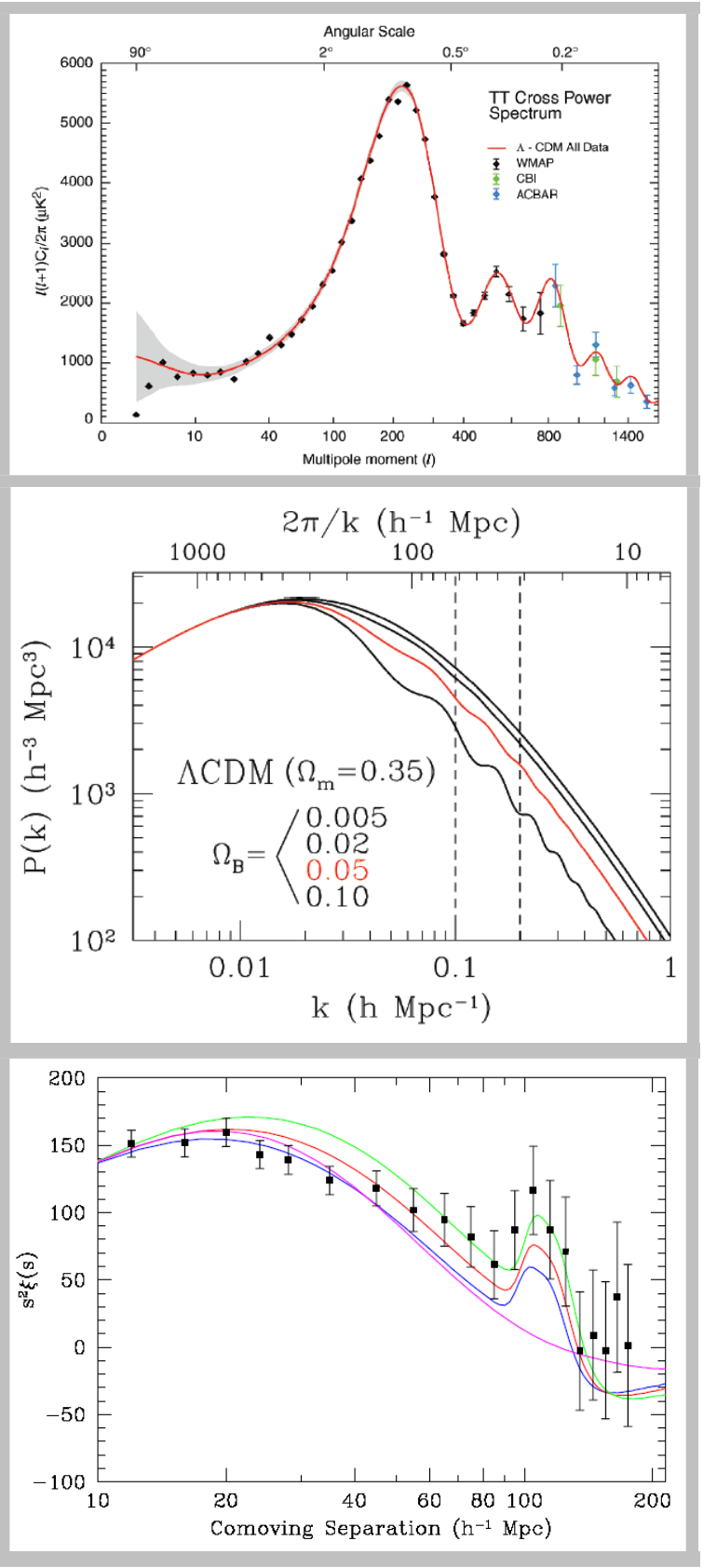

Fig 1: The acoustic oscillations as mapped by the WMAP first-year data, CBI and ACBAR data. Also shown is the best-fit ΛCDM model. In Fig. 2 we show the corresponding theoretical linear power spectra for dark matter as a function of the amount of baryons in the universe (controlled by $\Omega_b$). [WMAP team]

Fig 2: The predicted matter power spectrum for various Λ Cold Dark Matter (ΛCDM) models showing the impact of baryons on the spectrum with the amount of dark matter, $\Omega_m$, fixed. The top curve has essentially no baryons and therefore shows no baryon acoustic oscillations while increasing the baryon content increases the amplitude of the oscillations. The wavelength of the oscillations provides a standard ruler for measuring cosmic distances accurately (Eisenstein et al. 1998).

Fig 3: The correlation function of luminous red galaxies (LRGs) from the SDSS survey, multiplied by the square of the separation, s (Eisenstein et al. 2005). The peak at about 100 $h^{-1}$MPc is the acoustic baryon oscillation feature shown with various model fits to the data with different values of $\Omega_b h^2$. The curve without a peak has no baryons in it at all. Unlike the power spectrum, the correlation function exhibits a single peak (c.f. Fig. 2).

The first of these is the 2dF Galaxy Redshift Survey (2dFGRS)[1] [Cole *et al.*, 2005; Colless *et al.* 2001], which is an Anglo-Australian project, and has mapped the redshifts and positions of 220,000 galaxies in the southern hemisphere and which finished taking data in 2002. The other is the Sloan Digital Sky Survey (SDSS)[2] [York *et al.* 2000], which is designed to measure redshifts of about one million galaxies in the north, and will continue until 2008.

The detection of the acoustic oscillations in the CMB gave us an exquisite measurement of the curvature of the Universe because the position of the first peak is determined by the angular scale subtended by a single physical scale at decoupling – the sound horizon. In a non-flat cosmos, this scale would look bigger or smaller than BOOMERanG and WMAP see it to be. It is just about in the right place expected for a flat cosmos. With the detection of these acoustic oscillations in the distribution of galaxies in the local Universe we can extend this method even further. Combining the late and early observations now provides cosmologists with a way to fix how much the cosmos has expanded since the CMB was formed.

In fact, the SDSS team has already used this "standard ruler" method to confirm that the energy density of the Universe is roughly three-quarters dominated by dark energy, in excellent agreement with the SNIa observations [Eisenstein *et al.* 2005][3]. This is possible as we now have observations of this ruler at two different redshifts, namely z=0.35 (the mean redshift of the SDSS study) and z=1089 (the redshift of the CMB surface of last scattering). Using these two observations of this standard ruler, the SDSS determined the ratio of the distances between these two redshifts to 4% fractional accuracy and the absolute distance to z = 0.35 to 5% accuracy, namely 1370 +/- 69 Mpc.

The SDSS measurements above use the angular diameter distance, which suffers the same double integral over redshift as discussed above for the SNIa measurements (although we stress the two measurements are completely independent and highly complementary). However, the acoustic oscillations exist not only across the sky (as used by the SDSS measurement) but also radially along the line of sight in any given direction. Therefore, if one can accurately map the three dimensional positions of galaxies then these radial acoustic oscillations provide a wealth of additional information and give us *H(z)*, as the radial length ($\Delta l$) of a structure stretching over a redshift interval, $\Delta z$, is simply given by $\Delta l = \Delta z / H(z)$.

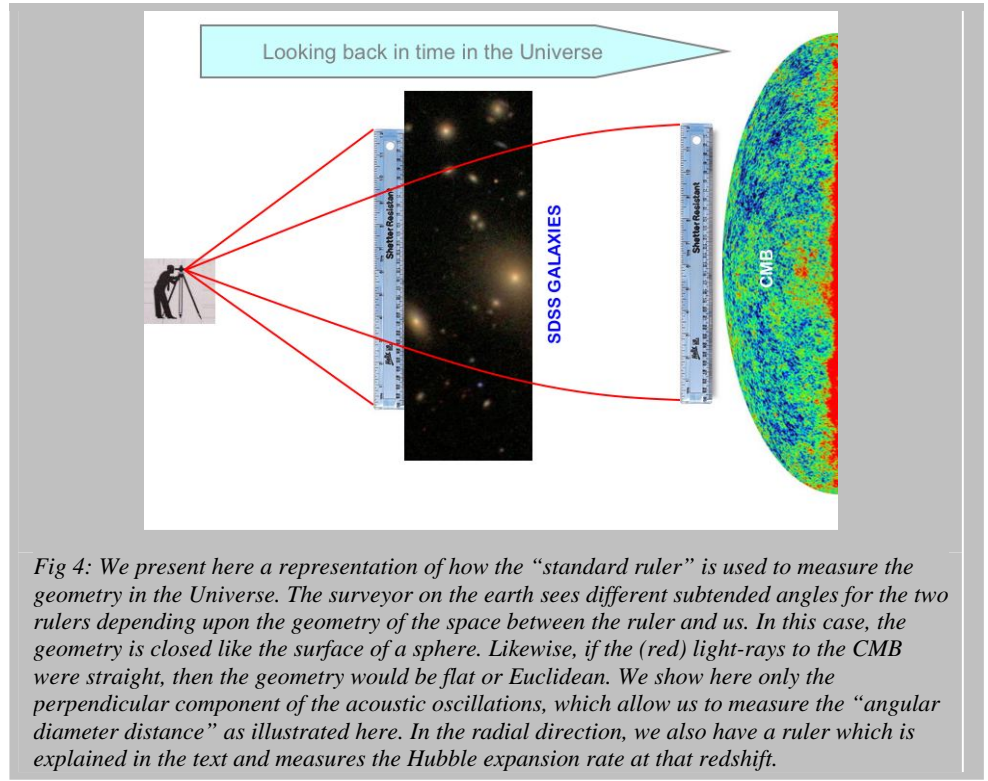

*Fig 4: We present here a representation of how the "standard ruler" is used to measure the geometry in the Universe. The surveyor on the earth sees different subtended angles for the two rulers depending upon the geometry of the space between the ruler and us. In this case, the geometry is closed like the surface of a sphere. Likewise, if the (red) light-rays to the CMB were straight, then the geometry would be flat or Euclidean. We show here only the perpendicular component of the acoustic oscillations, which allow us to measure the "angular diameter distance" as illustrated here. In the radial direction, we also have a ruler which is explained in the text and measures the Hubble expansion rate at that redshift.*

## Future Acoustic Oscillation Surveys

The key advantages of the baryon oscillation method therefore are that it provides both the distance to a given redshift, z, *and* the Hubble expansion rate at that redshift, *H(z)*. In addition, the physical origin of these oscillations is already well understood since it only involves linear physics! Controlling systematic errors is therefore significantly easier than it is for every other method which always involves understanding complex nonlinear physics, whether that be SNIa explosions, cluster formation or any other method so far proposed.

Given these significant advantages of the baryon oscillations (BAO) over other methods of mapping the Universe, and their recent discovery in the local Universe, there are now a number of teams clamoring to undertake new galaxy surveys focused on measuring the BAO as a function of redshift. This is a challenging prospect however, as to detect and measure the oscillations one must perform surveys with sufficient volume to prevent the oscillations being washed out by the convolution with the window function of the survey, and must contain hundreds of thousands of galaxies to sufficiently sample the underlying density field. To achieve these goals, most future surveys will need to probe the distant Universe (e.g. z>0.5) to gain the necessary volume aswell as providing important redshift leverage on dark energy models: This naturally implies that future BAO surveys will need to be performed on large telescopes with large fields-of-view to

---

gain the required volume, object densities and redshift coverage. Future BAO experiments can be broken down into three distinct classes. The first of these is the expectations for further BAO constraints from existing data and surveys. For example, the SDSS will continue to take data until 2008 and hopes to complete its original footprint of 10,000 square degrees of multi-color imaging data and over one.million galaxy redshifts over this area, including 100,000 Luminous Red Galaxies (LRGs), about half of which constituted the source of the original SDSS BAO discovery. The extra SDSS data will therefore only improve the existing BAO detections by about $\sqrt{2}$ and is only at low redshifts, $z \sim 0.35$.

The second class is the use of existing or planned large-area multi-color imaging surveys to detect the acoustic oscillations. These surveys will use "photometric redshifts", or crude estimates of the trueredshift of all detected galaxies based on their observed colours, to detect and measure the perpendicular component of the acoustic oscillations in the angular clustering of galaxies in a handful of broad redshift shells. Such measurements are already underway using just the SDSS imaging data, but the errors induced by the photometric redshifts need to be compensated by imaging huge areas of the sky. Other large-area imaging surveys that may provide angular detections of the BAO over the next decade include the UK-led UKIDSS and VISTA near infrared (IR) surveys, as well as the proposed darkCAM and Dark Energy Survey (DES) and the LSST in the optical, but none of these will yield good measurements of the Hubble expansion rate, $H(z)$, since one needs to map the radial positions of galaxies accurately to do this.

To recover the Hubble rate information, a massive new spectroscopic redshift survey of distant galaxies is required. Therefore, the third class of BAO experiments will focus on constructing such surveys over the next ten years.In the short-term, new facilities like the AAOmega spectrographs on the Anglo-Australia Telescope and FMOS on the Subaru Telescope will provide a first opportunity to achieve this goal and, if given significant observing time, could obtain hundreds of thousands of galaxy redshifts over large areas of sky. It is possible that within the next 5 years, these new instruments will measure the BAOwith the same accuracy as the SDSS and 2dFGRS detections but at higher redshifts, $z > 0.5$.

On a somewhat longer time-scale (2012), an international team of astronomers from the UK, US, Australia, Canada and Japan are proposing to construct a major new facility for the Gemini & Subaru observatories called the "Wide Field Multi-Object Spectrograph" (WFMOS). The WFMOS instrument has developed from the initial Kilo-Aperture Optical Spectrograph (KAOS)[4] concept and will represent a quantum leap in our present efficiency of collecting spectra of astronomical objects; observing up to five thousand sources at a time.

The science drivers for WFMOS/KAOS are high precision measurements of the acoustic oscillations at high redshift and the study of the formation history of our Galaxy. Both of these science goals demand massive spectroscopic surveys, e.g. combined spectra for over two and a half million galaxies at $z\sim1$ and $z\sim3$, over 2000 square degrees of sky, for measurement of the acoustic oscillations and over half a million stars for the determination of the chemical abundance and dynamical history of our Galaxy. This is only possible with new instrumentation on 8-meter telescopes. WFMOS/KAOS will also obtain spectra for hundreds of thousands of stars in our Galaxy to study their motions and element abundances and thus unravel the merger history of our Galaxy.

WFMOS has recently undergone a detailed Feasibility Study, which has investigated all technical and scientific aspects of this major new project. In figure 6 we show the latest predictions of the likely accuracy of the angular diameter distance and Hubble rate from the proposed WFMOS measurements of the acoustic oscillations. As one can see, WFMOS will obtain 2% measurements of both these quantities by a projected date of 2013. The present baseline design of WFMOS is to provide 4500 fibres over a 1.5 degree field-of-view on a 8-meter class telescope, which will be achieved using an Echidna fiber-positioner feeding 10 low resolution SDSS-type spectrographs to a resolution of R~4000 and 2 high resolution spectrographs each handling approximately 750 fibres to R~40,000, shown schematically in Figure 5 in a possible configuration for the Subaru telescope.

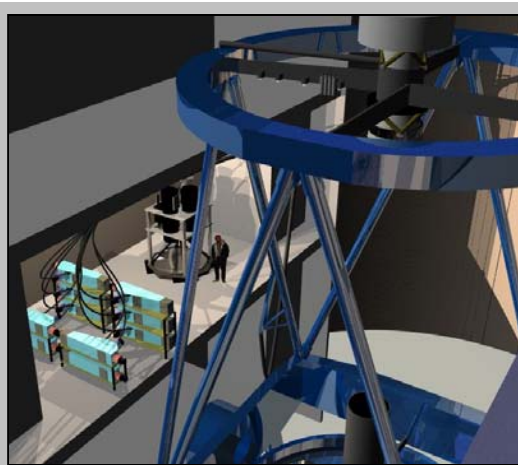

*Fig.5: A WFMOS spectrograph room in the Subaru telescope dome, with 10 spectrographs based on Johns Hopkins University's SDSS design to accept a total of 3000 fibres, and a double-spectrograph unit producing 1500 high-resolution spectra. The 1.5-degree wide field corrector, fibre positioner housing and other WFMOS components can be seen at the top of the telescope. The existing wall alongside the telescope has been omitted to show the spectrographs. [WFMOS Feasibility Study]*

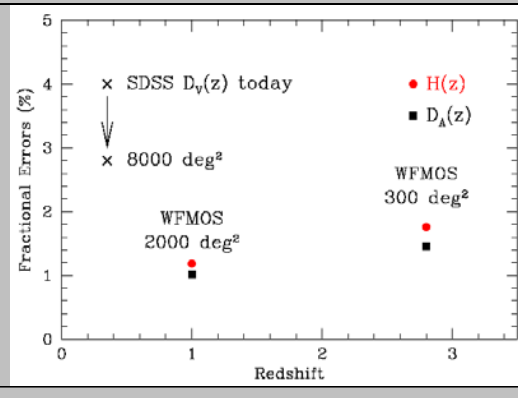

*Fig 6: Expected errors on the distance and Hubble rate, H(z), as a function of redshift for the two fiducial WFMOS/KAOS surveys at high and low redshifts together with the current and future constraints from the SDSS survey at z=0.35. By way of comparison the Hubble Space Telescope key project result only gives the Hubble rate at z=0 accurate to about 10%. [WFMOS Feasibility Study]*

---

[4] http://www.dsg.port.ac.uk/~bruce/kaos

The WFMOS will be a unique instrument and represents the next generation of spectroscopic facilities. In addition to performing dedicated surveys to measure dark energy and the history of our Galaxy, WFMOS will be a facility instrument available for proposal-based projects like the present 2dF/AAOmega and FMOS instruments. Such community-driven science will provide a wide variety of science, e.g., the study of the large scale structure of the Universe through mapping absorption lines in distant quasars. Furthermore all these data (surveys and community proposals) will stimulate a host of archival science including the astrophysics of galaxy evolution and new measurements on the mass of neutrinos and possible interactions with axions – the proposed dark matter particle that fills our Universe. In summary, WFMOS will simultaneously provide a window on our Galaxy, the distant Universe and fundamental particles.

The WFMOS dark energy survey would comprise two parts. The first would be a large (2000 square degree) survey of about two million galaxies at $z < 1.3$ (at higher redshifts optical galaxy emission is received in the infra red). This would be complemented by a high redshift survey of about half a million Lyman Break Galaxies (LBGs) which can be found readily at redshifts $2.5 < z < 3.5$. LBGs are good tracers of large-scale structure (they were the objects used in the recent SDSS detection of baryon oscillations) and are easily detected as they are young, starbursting galaxies which makes them easy to find because they have blue spectra longward of the Lyman break (912Å). Shortwards of the break all light is absorbed by neutral HI at $2.5<z<3.5$. They are thus selected by very red $U$–$B$ colors combined with very blue $B$–$R$ colors.

The LBG part of the WFMOS survey would provide the first direct probes of distance and Hubble rate beyond $z = 2$ and, if past experience is a guide, may well have surprises in store for us.

Up until now we have focused almost exclusively on the WFMOS capability to track dark energy. This is of course just a small part of the cosmology one can undertake with this survey which would also provide us with completely new maps of large-scale structure and clustering at high redshifts. This would allow us to study the nature of clustering at high redshift and to watch the evolution of non-Gaussianity, biasing and redshift distortions.

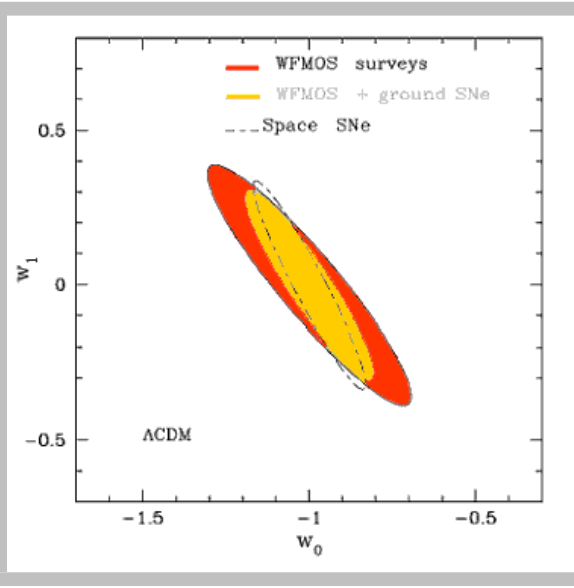

Fig. 7: projected error ellipses on the dark energy equation of state, $w(z) = w_0 + w_1 z$, where w is the ratio of dark energy pressure to dark energy density for WFMOS/KAOS alone and with ground-based SNIa data and compared with projected errors for the space-based SNAP SNIa satellite scheduled for launch after 2014. [WFMOS Feasibility Study]

The high-redshift LBG part of the survey has a further advantage – it is linear on much smaller scales than it is today where non-linearity smears out the oscillations we are chasing. As a result we will be able to study the shape of the spectrum down to small scales and look for the characteristic deviations due to neutrino masses and running of the spectral index predicted by some models of inflation.

Conservative estimates are that WFMOS will be able to weigh neutrinos to an accuracy of less than 0.2 eV [Lesgourgues *et al*. 2004]. If the sum of the three neutrino masses is larger than this WFMOS will detect the neutrino masses which may provide crucial insights into physics beyond the standard model of particle physics.

## Summary

The power spectrum of galaxies may give us insights into physics at much higher energies too. The final power spectrum we see in the late universe is the product of the primordial spectrum – laid down in the first second of the history of the cosmos in standard models – and a transfer function, encoding the subsequent fourteen billion years free-falling under the influence of gravity. Assuming we understand this transfer function, WFMOS's observations – together with the Planck satellite scheduled for launch in 2008 – will allow us to constrain models of inflation.

Most inflationary models predict that the primordial spectrum of fluctuations should be very close to Harrison-Zel'dovich type, that is, scale-invariance with constant power in every log k interval.

However, the SDSS and WMAP data give tantalizing evidence for departures from scale-invariance. Confirmation of these scale-dependent features would give us a powerful window into physics happening within the first fraction of a second after the Big Bang. WFMOS may turn up new surprises about the cosmos at early and late times, and may lead to fundamental breakthroughs in our understanding of the physics of cosmic acceleration.

## Acknowledgements

The material presented here is partly based on the WFMOS/KAOS feasibility study funded by Gemini. We thank all other members of the WFMOS/KAOS feasibility study team for useful discussions. We thank Roy Maartens and Andrew McGrath for detailed comments on this article and Andrew McGrath for supplying Figure 5.